\pdfoutput=1

\input{aipcheck}

\documentclass[
    ,final            
  ]
  {aipproc}

\layoutstyle{8x11single}

\begin{document}

\title{Results on Charmonium(-like) and Bottomonium(-like) States from Belle and BaBar}


\author{Jens S\"oren Lange, for the Belle Collaboration}{
  address={Justus-Liebig-Universit\"at Gie\ss{}en, II.\ Physikalisches Institut, 
  Heinrich-Buff-Ring 16, 35392 Gie\ss{}en, Germany}
}


\begin{abstract}
Spectroscopy results for Belle and BaBar are reported.
A particular focus is put on new results of the X(3872) state
with its radiative decays to $J$/$\psi$$\gamma$ and $\psi'$$\gamma$,
its decay into $J$/$\psi$3$\pi$ and 
the search for production in radiative Upsilon decays.
Another focus is $L$=2 mesons,
in particlar a possible $D$-wave assignment to the X(3872) 
and the confirmation of an Upsilon $D$-wave state.
\end{abstract}

\maketitle

\noindent
Conference Talk presented at MENU10, 12$^{th}$ International Conference on Meson-Nucleon Physics 
and the Structure of the Nucleon, 05/31-06/04, 2010, College of William and Mary, Williamsburg, Virginia.

\section{Introduction}

\noindent
In this paper, spectroscopy results 
from the two B factories 
BaBar \cite{babar_detector} at PEP-II and 
from Belle \cite{belle_detector} at KEKB 
are reported.
The data samples of the two experiments 
are summarized in Tab.~\ref{tlumi}.

\noindent
There is quite a number of unresolved interesting questions 
in charmonium spectroscopy, such as: 

\begin{itemize}

\item The charmonium potential is usually regarded as a static
quark-antiquark potential (Cornell ansatz) \cite{potential_1978}

\begin{equation}
V(r) = - \frac{4}{3} \frac{\alpha_S}{r} + k \cdot r
\end{equation}

with a Coulomb-like term and a confinement term.
As can be seen, $\alpha_S$ is assumed fixed over the total
range 0$<$$r$$<$$\simeq$1~fm, which is an approximation.
Also, as the mechanism of confinement is still one of the unanswered questions
in QCD, it is still unproven if the 
linear approximation (corresponding to a constant string force)
in the confinement term is {\it (a)} valid to all orders of $\alpha_S$
and {\it (b)} valid even in the far long range $r$$>$$\simeq$1.3~fm 
(i.e.\ in the string breaking regime).
In fact, studies of potential constructed with two-gluon exchanges 
\cite{potential_2gluon_1} \cite{potential_2gluon_2}
lead to a number of additional terms with different $r$ dependances.
In addition, the high lying states (e.g.\ $L$$\leq$2, $n$$\geq$3) are sensitive 
to the string constant $k$$\simeq$1~GeV/fm, which is the 
slope of the confinement term, and mass measurements 
can provide a precise measurement of $k$. 

\item The strong coupling constant in the charmonium system 
has a quite high value $\alpha_S$=0.54 \cite{potential_2005}, so there might be 
non-perturbative effects becoming visible e.g.\ in the hyperfine splittings.
 
\item What is the nature of the newly observed narrow states near thresholds,
which do not fit into potential model calculations?
Are they molecular states? Or tetraquarks? Or threshold effects?
As an example, molecular potentails might contain $r^{-2}$ and $r^{-3}$ 
\cite{potential_molecular}
terms, not present in the Cornell potential, and leading to eigenstates
with masses different from the quark-antiquark potential.

\item There are yet unobserved states, some of them expected narrow,
e.g.\ the $^3D_2$(J$^{PC}$=2$^{--}$) state.

\end{itemize}

\noindent
Similar questions apply to the bottomonium system.
At $B$ factories, studies of bottomonium require the change of the beam energies.
A few results are reported down below.

\section{Width of $\eta_c$}

\noindent
Although the $\eta_c$ is the ground state of the charmonium system 
(J$^{PC}$=0$^{-+}$, $^1S_0$) and already discovered in 1980 
\cite{etac_1980_mark2} \cite{etac_1980_cb}, 
its width has been of particular interest recently.
Previous width measurements \cite{pdg} showed values around 
$\Gamma$$\simeq$15~MeV from radiative $J$/$\psi$ and $\psi'$ decays, 
and values around $\Gamma$$\simeq$30~MeV from $B$ meson decays.
However, this is probably not surprising, as in the radiative decays 
the cross section varies according to $E_{\gamma}^a$ with an exponent $a$=3$-$7.
This energy dependance modifies the lineshape
and the width determination becomes non-trivial.
On the other hand, in the reaction $\gamma$$\gamma$$\rightarrow$$\eta_c$
the Breit-Wigner line shape is an appropriate approximation.
In a new high statistics measurement by BaBar 
with a data set of 469~fb$^{-1}$ and 14090 $\eta_c$ signal events,
a high precision measurement of the mass 
$m$=2982.2$\pm$0.4$\pm$1.6~MeV
and the width $\Gamma$=31.7$\pm$1.2$\pm$0.8~MeV
of the $\eta\_c$ could be obtained.
This measurement represents a factor $\simeq$3 improvement 
in both statistical and systematic errors compared to 
the BaBar measurement in 2008 in $B$ meson decays \cite{etac_2008_babar} and
the Belle measurement in 2008 in $\gamma$$\gamma$ collisions \cite{etac_2008_belle}.

\begin{table}
\begin{tabular}{lrr}
\hline
  & \tablehead{1}{c}{b}{Belle} & \tablehead{1}{c}{b}{BaBar}\\
\hline
On-resonance & & \\
\hline
$\Upsilon$(4S) & 711 fb$^{-1}$ & 433 fb$^{-1}$ \\
$\Upsilon$(5S) & 121 fb$^{-1}$ & $-$ fb$^{-1}$ \\
$\Upsilon$(3S) & 3.0 fb$^{-1}$ & 30 fb$^{-1}$ \\
$\Upsilon$(2S) & 24 fb$^{-1}$ & 14 fb$^{-1}$ \\
$\Upsilon$(1S) & 5.7 fb$^{-1}$ & $-$ fb$^{-1}$ \\ 
\hline
Off-resonance & & \\
\hline
 & 87 fb$^{-1}$ & 54 fb$^{-1}$ \\
\hline
Total & 952 fb$^{-1}$ & 553 fb$^{-1}$ \\
\hline
\end{tabular}
\caption{Integrated luminosities for data sets at different beam energies for BaBar and Belle.}
\label{tlumi}
\end{table}

\subsection{Decays of X(3872) to $D$$\overline{D}^*$}

\noindent
The X(3872) state has been discovered in $B$ meson decays
in the decay X(3872)$\rightarrow$$J$/$\psi$$\pi^+$$\pi^-$
by Belle \cite{x3872_belle}. 
It was confirmed in the same process by BaBar \cite{x3872_babar} 
and confirmed in inclusive production in $p$$\overline{p}$ production
at $\sqrt{s}$=1.8~TeV at CDF-II \cite{x3872_cdf} and D0 \cite{x3872_d0}.
Among the newly observed charmonium-like states (sometimes referred to as 
XYZ states) the X(3872) is the only one with several decay channels
having been observed.
It has a surprisingly very narrow width $\Gamma$<2.3 MeV \cite{x3872_belle}, 
although its mass is above the open charm threshold.
As its mass with 3871.56$\pm$0.22~MeV \cite{pdg} 
is very close (within 1~MeV) to the $D^0$$D^{0*}$
threshold, it was discussed as a possible
$S$-wave $[$$D^0$$D^{0*}$$]$ molecule \cite{tornqvist_1} \cite{tornqvist_2}.

\noindent
The decay into $D^{\pm}$$D^{\mp*}$ is kinematically forbidden, 
but the decay into $D^0$$D^{0*}$ is a strong decay and 
among the so far observed decays it represents the dominant one, 
i.e.\ the branching fraction is a factor $\simeq$9
higher than for the decay into $J$/$\psi$$\pi^+$$\pi^-$.
In this decay channel, BaBar 
measured surprisingly a high mass 
of the X(3872) as
$m$=3875.1$^{+0.7}_{-0.5}$(stat.)$\pm$0.5(syst.)~MeV 
\cite{ddpi_babar}.
This high value initiated discussion, that there might be two different X states,
namely X(3872) and X(3875), which would fit to a tetraquark hypothesis \cite{maiani}
and the two different states $[$$c$$\overline{c}$$u$$\overline{u}$$]$
and $[$$c$$\overline{c}$$d$$\overline{d}$$]$.
On the other hand Belle measured in the same decay channel the mass as 
$m$=3872.9$^{+0.6}_{-0.4}$(stat.)$^{+0.4}_{-0.5}$(syst.)~MeV \cite{ddpi_belle}
and thus consistent with the world average \cite{pdg}.
A possible explanation of the discrepancy is the 
difficulty of performing fits to signals close to threshold.
In fact, the two fits used two very different approaches:

\begin{itemize}

\item Babar used a 1-dimensional binned maximum likelihood fit \cite{ddpi_babar}
with the $D^*$$D$ invariant mass as the only variable,  
where the signal probability density function was extracted 
from MC simulations and an exponential function was used 
for the background parametrization.

\item Belle used an 2-dimensional unbinned maximum likelihood fit \cite{ddpi_belle}, 
i.e.\ on the one hand the beam constraint mass with a Gaussian signal 
and an Argus function for the background, and 
on the other hand a Breit-Wigner signal for the $D^*$$D$ invariant mass
with a square root function for the background. 

\end{itemize}

\subsection{Radiative Decays of X(3872)}

\noindent
The branching fraction of the rare decay X(3872)$\rightarrow$$J$/$\psi$$\gamma$ 
\hspace*{1mm} is 
a factor $\simeq$6 smaller than the one for X(3872)$\rightarrow$$J$/$\psi$$\pi^+$$\pi^-$.
The first evidence by Belle \cite{x3872_jpsigamma_belle} 
was based upon a data set of 256~fb$^{-1}$
with 13.6$\pm$4.4 signal events.
The signal was confirmed by BaBar \cite{x3872_jpsigamma_babar}
with a data set of 424~fb$^{-1}$ and 23.0$\pm$6.4 signal events.
Although rare, this decay channel is very important, as 
their observation clearly establishes a $C$=+1 charge parity 
assignment to the X(3872).

\noindent
Recently BaBar found evidence for the decay 
X(3872)$\rightarrow$$\psi'$$\gamma$ \cite{x3872_psiprimegamma_babar}
with 424~fb$^{-1}$ and 25.4$\pm$7.4 signal events. 
The signal yield of this observation was surprising, 
as it implied a large ratio 
BR(X(3872)$\rightarrow$$\psi'$$\gamma$)/
BR(X(3872)$\rightarrow$$J$/$\psi$$\gamma$)=3.4$\pm$1.4.
There was a priori no understanding of the fact, 
why the transition of the X(3872) to a $n$=2 charmonium state 
should be stronger than to $n$=1.
In fact, quite the opposite behaviour was expected.

\noindent
In case of X(3872)$\rightarrow$$J$/$\psi$$\gamma$
the photon energy is $E_{\gamma}$=775~MeV, and thus 
due to vector meson dominance $\rho$ and $\omega$ can contribute
to the amplitudes. However, in case of 
X(3872)$\rightarrow$$\psi'$$\gamma$
with the smaller $E_{\gamma}$=186~MeV
the transition can only proceed through light quark 
annihilation with an expected small amplitude.
A new measurement by Belle of both radiative channels 
was based upon a data set of 711~fb$^{-1}$ \cite{x3872_psiprimegamma_belle}.
The background was studied in MC simulations
and revealed peaking behaviour in some background components
close to the signal region.
The signal X(3872)$\rightarrow$$J$/$\psi$$\gamma$ was clearly reestablished
with 
30.0$^{+8.2}_{-7.4}$ signal events 
(4.9$\sigma$ significance)
for $B^+$$\rightarrow$$K^+$X(3872)
and 
5.7$^{+3.5}_{-2.8}$ signal events  
(2.4$\sigma$ significance)
for $B^0$$\rightarrow$$K^0$X(3872).

\begin{figure}[htb]
\unitlength1cm
\begin{picture}(18,9)
\put(0.7,0){\includegraphics[width=\textwidth,bb=0 0 1221 593]{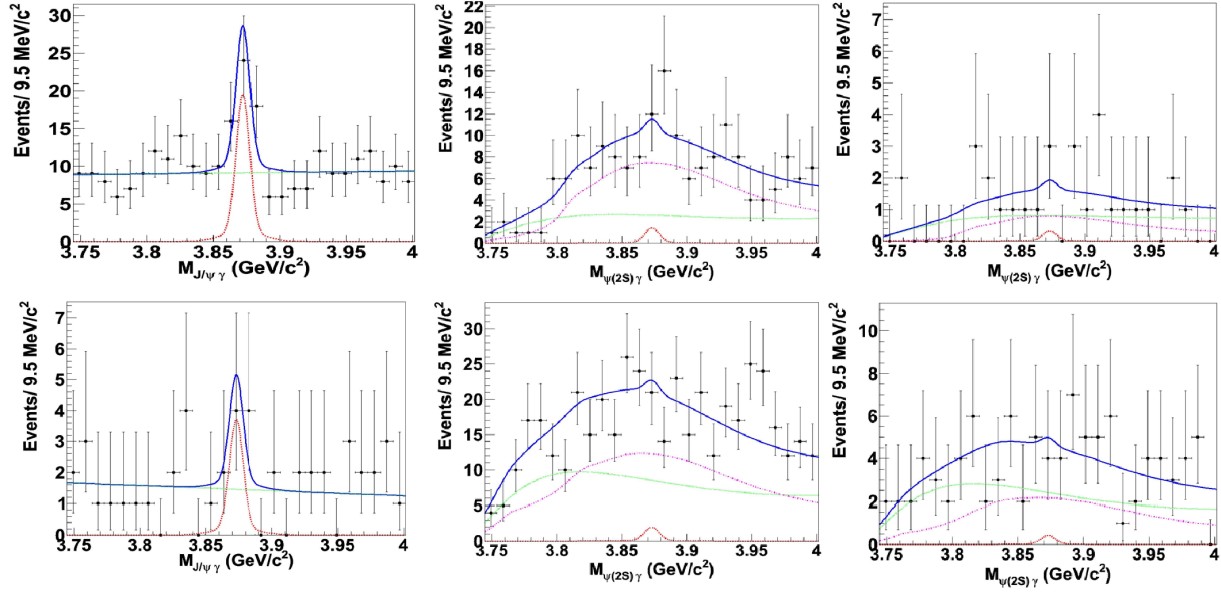}}
\end{picture}
\caption{Preliminary Belle results on radiative decays of the X(3872).
$J$/$\psi$$\gamma$ invariant mass in the X(3872) mass region
for 
$B^+$$\rightarrow$$K^+$X(3872)($\rightarrow$$J$/$\psi$$\gamma$) (top left)
and $B^0$$\rightarrow$$K_s^0$X(3872)($\rightarrow$$J$/$\psi$$\gamma$) (bottom left), 
and 
$\psi'$$\gamma$ invariant mass in the X(3872) mass region
for 
$B^+$$\rightarrow$$K^+$X(3872)($\rightarrow$$\psi'$$\gamma$) (top center)
$B^0$$\rightarrow$$K_s^0$X(3872)($\rightarrow$$\psi'$$\gamma$) (bottom center)
with $\psi'$$\rightarrow$$e^+$$e^-$,$\mu^+$$\mu^-$ 
and 
$B^+$$\rightarrow$$K^+$X(3872)($\rightarrow$$\psi'$$\gamma$) (top right)
$B^0$$\rightarrow$$K_s^0$X(3872)($\rightarrow$$\psi'$$\gamma$) (bottom right)
with $\psi'$$\rightarrow$$J$/$\psi$$\pi^+$$\pi^-$. 
The solid, dotted and dot-dashed curves are the total, 
combinatorial background and combined $\psi'$$K^*$ and $\psi'$$K$ background, 
respectively. The background subtracted signal is also shown. 
}
\end{figure}

\noindent
For X(3872)$\rightarrow$$\psi'$$\gamma$, the four decay channels
$B^+$$\rightarrow$$K^+$X(3872) and $B^0$$\rightarrow$$K^0$X(3872)
with $X$(3872)$\rightarrow$$\psi'$$\gamma$ with 
$\psi'$$\rightarrow$$l^+$$l^-$ and 
$\psi'$$\rightarrow$$J$/$\psi$$\pi^+$$\pi^-$.
Charged and neutral $B$ modes were treated separately, 
but the two $\psi'$ subdecay modes 
were fitted simultaneously because of their different
background shapes. The signal was treated as a double
Gaussian, the combinatorial background was parameterized 
as a threshold function.
The shape of the $\psi' K^*$ and $\psi' K$ background,
and in particular the peaking structures, 
was modeled as a sum of bifurcated gaussians
using a large MC sample. 
The signal yields were determined as
5.0$^{+11.9}_{-11.0}$ signal events 
(0.4$\sigma$ significance)
for $B^+$$\rightarrow$$K^+$X(3872)
and 
1.5$^{+4.8}_{-3.9}$ signal events  
(0.2$\sigma$ significance)
for $B^0$$\rightarrow$$K^0$X(3872).
Thus, contrary to BaBar, Belle observed no signal, which 
would imply that there is no indication that the 
radiative transition from X(3872) to $n$=2 charmonium 
is stronger than to $n$=1 charmonium.

\noindent
In the same analysis, 
the decay $\chi_{c1,2}$$\rightarrow$$J$/$\psi$$\gamma$ 
was used as a reference channel
with signal yields of 
32.8$^{+10.9}_{-10.2}$
(3.6$\sigma$ significance)
for $B^+$$\rightarrow$$K^+$$\chi_{c2}$
and 
2.8$^{+4.7}_{-3.9}$
(0.7$\sigma$ significance)
for $B^0$$\rightarrow$$K^0$$\chi_{c2}$.
In the charged mode, this represents the first observation
of a $J^P$=$2^+$ state 
in a rare exclusive final state in a $B$ meson decay,
and thus a transition $0^-$$\rightarrow$$0^-$$2^+$.

\subsection{Decays of X(3872) and Y(3940) into $J$/$\psi$3$\pi$}

\noindent
As the $\rho$ meson carries isospin $I$=1, 
the X(3872) seems to violate isospin conservation in the decay 
X(3872)$\rightarrow$$J$/$\psi$$\rho$($\rightarrow$$\pi^+$$\pi^-$).
One of the proposed explanations \cite{terasaki} 
is $\rho$/$\omega$ mixing, and therefore the 
investigation of the decay 
X(3872)$\rightarrow$$J$/$\psi$$\omega$($\rightarrow$$\pi^+$$\pi^-$$\pi^0$)
is of importance.
The difficulty hereby is the nearby Y(3940) state, 
which is also known to decay into the same final state.
The latter decay was investigated be Belle 
with a data set of 275$\times$10$^6$ $B$ meson pairs.
The mass of the Y(3940) 
was determined as 3943$\pm$11(stat.)$\pm$13(syst.)~MeV
with a width of 87$\pm$22(stat.)$\pm$26(syst.)~MeV.
Belle also observed a signal 
for X(3872)$\rightarrow$$J$/$\psi$$\omega$($\rightarrow$$\pi^+$$\pi^-$$\pi^0$),
based upon a data set of 256~fb$^{-1}$ \cite{y38723pi_belle}.
The measured efficiency corrected ratio of 
X(3872)$\rightarrow$$J$/$\psi$$\pi^+$$\pi^-$$\pi^0$/
X(3872)$\rightarrow$$J$/$\psi$$\pi^+$$\pi^-$=
1.0$\pm$0.4(stat.)$\pm$0.3(syst.) indicated 
another case of isospin violation due to the additional $\pi^0$.
BaBar published slightly different values for the Y(3940)
with a data set of 383$\times$10$^6$ $B$ meson pairs mass \cite{y3940_babar}, 
namely a mass of 3914.6$^{+3.8}_{-3.4}$(stat.)$\pm$2.0(syst.)~MeV
and a width of 34$^{+12}_{-8}$(stat.)$\pm$5(syst.)~MeV.
However, a signal for X(3872)$\rightarrow$$J$/$\psi$$\omega$($\rightarrow$$\pi^+$$\pi^-$$\pi^0$)
was not observed.
In a recent re-analysis with 433~fb$^{-1}$ by BaBar, 
a requirement on the 3-pion mass was adjusted,
i.e.\ the lower offset was extended from 0.7695~GeV to 0.7400~GeV.
With this change in the analysis technique 
BaBar was able to confirm the Belle signal for
X(3872)$\rightarrow$$J$/$\psi$$\omega$($\rightarrow$$\pi^+$$\pi^-$$\pi^0$)
and confirm the large isospin violation for the ratio
X(3872)$\rightarrow$$J$/$\psi$$\pi^+$$\pi^-$$\pi^0$/
X(3872)$\rightarrow$$J$/$\psi$$\pi^+$$\pi^-$ measured
as 0.7$\pm$0.3(stat.) and 1.7$\pm$1.3(stat.) 
for $B^+$ and $B^0$ decays, respectively.

\noindent
In the re-analysis, BaBar also investigated the shape of the 3$\pi$ mass distribution
in order to determine the quantum number of the X(3872).
A similar analysis for the 2$\pi$ mass distribution in case of 
X(3872)$\rightarrow$$J$/$\psi$$\pi^+$$\pi^-$
was performed before by Belle \cite{x38722pi_belle} 
and CDF-II \cite{x38722pi_cdf}. The result in both cases
was that $S$-wave is preferred.
However, in the new BaBar analysis the shape of the 3$\pi$ mass
distribution seems to indicate that $P$-wave is preferred.
For the 2$\pi$ case and $S$-wave, a parity of $+1$ for the X(3872) is preferred,
leading to a tentative assignment of $J^{PC}$=1$^{++}$.
This quantum number assignment is also supported by angular analyses
\cite{y38723pi_belle} \cite{x3872angular_cdf} and leads 
to a possible charmonium state assignment of
$\chi'_{c1}$ ($^3P_1$), which is an $n$=2 state
with a mass of 3953~MeV as predicted by potential models \cite{potential_2005}
and thus $\simeq$70 MeV too high compared to the observation.
For the 3$\pi$ case and $P$-wave, a parity of $-1$ for the X(3872) is preferred,
leading to a possible assignment of $J^{PC}$=2$^{-+}$.
Then a possible charmonium assignment is
$\eta_{c2}$ ($^1D_2$), which is an $n$=1 state.
The predicted mass is $\simeq$100 MeV lower than for the $\chi'_{c1}$.
This state would be an $L$=2 meson.

\noindent
In a different analysis, Belle investigated the $J$/$\psi$$\omega$ final state
in $\gamma$$\gamma$ collisions based upon 694~fb$^{-1}$ \cite{jpsiomega_gammagamma_belle}. 
This analysis not only uses $\Upsilon$(4S), but also $\Upsilon$(3S) and $\Upsilon$(5S) data (see Tab.~\ref{tlumi}).
The event selection uses a $p_T$$<$0.1~GeV/c balance requirement.
The final state in $\gamma$$\gamma$ collisions is required to have isospin $I$=0.
Fig.~2 shows the 
$W$ distribution of the final candidate events, 
where $W$ is defined as $W$=$m_5$-$m$($l^+$$l^-$)+$m_{J/\psi}$.
$m_5$ is the invariant mass of the system constructed
from four charged tracks and a neutral pion candidate.
A clear enhancement seen just above $J$/$\psi$$\omega$ threshold
with 49$\pm$14(stat.)$\pm$4(syst.) events (7.7$\sigma$ stat.\ significance).
The fitted mass is 3915$\pm$3(stat.)$\pm$2(syst.) MeV,
thus it might be the observation the Y(3940) state
in a second production mode.
However, the fitted width is $\Gamma$=17$\pm$10(stat.)$\pm$3(syst.)~MeV,
which is narrower than the width of the Y(3940) as measured in $B$ decays.
The production mode allows to establish the charge parity $C$=$+$1 for this state, 
same as the X(3872), 
but the determination of the other quantum numbers would require more statistics.

\begin{figure}[htb]
\unitlength1cm
\begin{picture}(17,5)
\centerline{\includegraphics[height=.23\textheight, width=7cm]{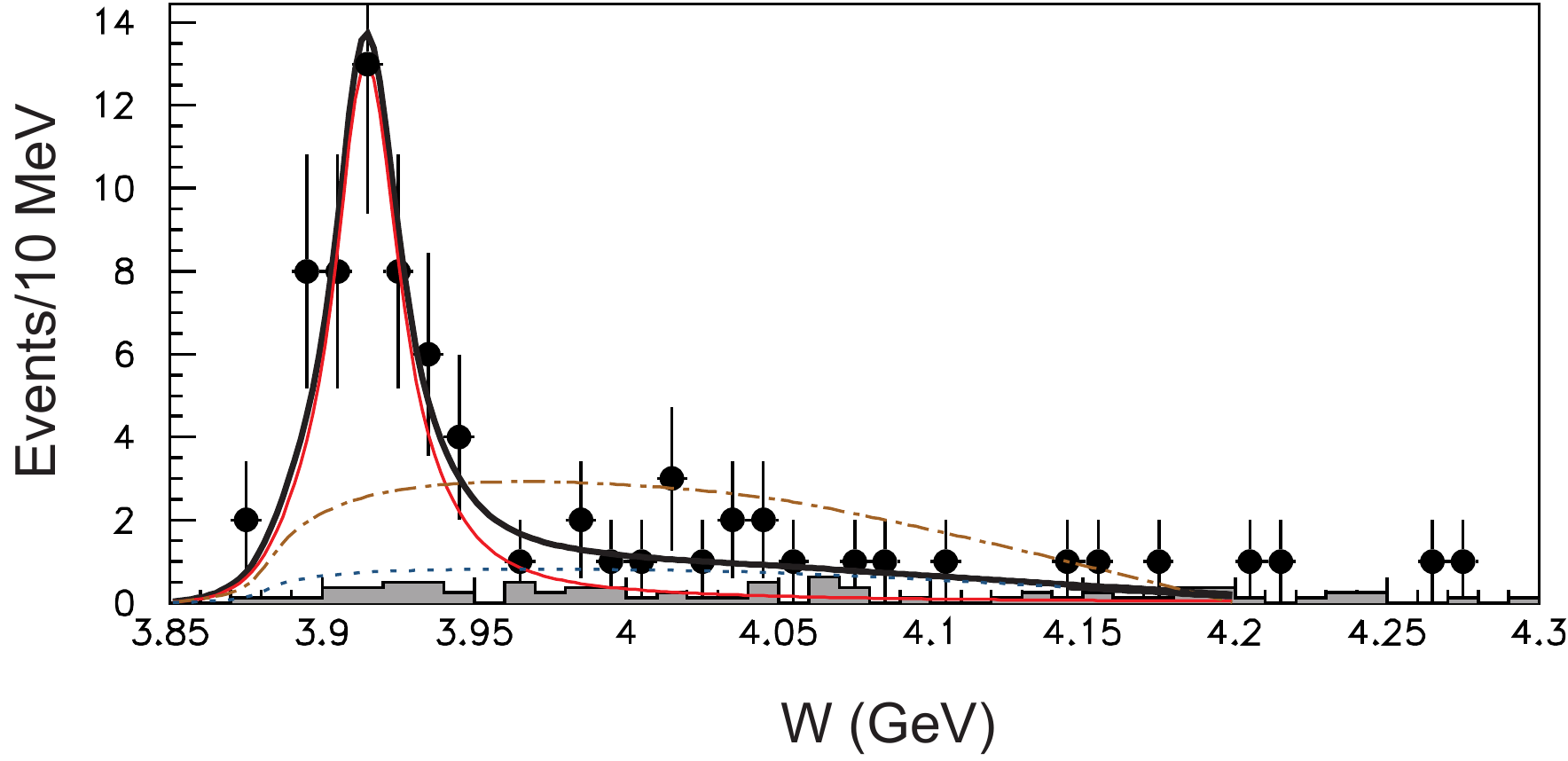}}
\end{picture}
\caption{The $W$ distribution of the final candidate events (dots
with error bars) for $\gamma$$\gamma$$\rightarrow$$J$/$\psi$$\omega$ at Belle 
\cite{jpsiomega_gammagamma_belle}.
The shaded histogram is the distribution of non-$J$/$\psi$ background
estimated from the sideband distribution. The bold solid, thinner solid 
and dashed curves are the total, resonance and background contributions, 
respectively. The dot-dashed curve is the fit without a resonance.
}
\end{figure}

\subsection{$\Upsilon$(1S) Radiative Decays to X(3872)}

\noindent
As shown in Tab.~\ref{tlumi}, Belle recorded an extensive data set 
with the beam energies adjusted to the $\Upsilon$(1S) resonance, 
the $n$=1~$^3S_1$ $b$$\overline{b}$ state with $J^P$=1$^-$
and a mass of 9.46~GeV.
With this data set, radiative transistions 
$b$$\overline{b}$$\rightarrow$$c$$\overline{c}$$\gamma$
can be investigated.
These rare events with an expected branching fraction
of $\leq$10$^{-5}$ \cite{bbccgamma_theory} with 
interfering QED and QCD amplitudes.
The transition may be from a $1^{--}$ state, such as the $\Upsilon$(1S),
to a $1^{++}$ state, which is one of the most probably quantum
number assignments for the X(3872).

\noindent
Belle searched for the process
$\Upsilon$(1S)$\rightarrow$$\gamma$X(3872)($\rightarrow$$J$/$\psi$$\pi^+$$\pi^-$) 
with a data set of 5.712~fb$^{-1}$ \cite{bbccgamma_belle}, 
corresponding to 88$\times$10$^6$ $\Upsilon$(1S) decays
The photon detection required $E_{\gamma}^{lab}$>3.5~GeV and 
the recoil mass on four charged tracks being consistent with zero,
i.e.\ $-$2<$m_{recoil}$<2~GeV$^2$.
Initial state radiation (ISR) was treated in two different ways.
On the one hand, ISR events were rejected by a criterium
on the cms polar angle of the photon, i.e.\ $|$cos$\vartheta_{\gamma}^*$$|$$<$0.9
On the other end, ISR events for $\psi'$ production, with the same $J$/$\psi$$\pi^+$$\pi^-$ 
final state as the X(3872), were used as a crosscheck, and 
the cross section for this ISR process was determined as 20.2$\pm$1.1(stat.)~pb.
For the X(3872) one event in the signal region was observed, 
resulting in an upper limit for the product branching fraction
BR(Y(1S)$\rightarrow$$\gamma$X(3872)$\times$BR(X(3872)$\rightarrow$$J$/$\psi$$\pi^+$$\pi^-$)<2.2$\times$10$^{-6}$
at 90\% CL.

\section{The $\Upsilon$(1D) state}

\noindent
$D$-wave mesons with orbital angular momentum $L$=2 are
quite different, if we compare the charmonium and the bottonium system.
For the charmonium system, there are two observed D-wave states {\it above} the $D$$\overline{D}$
threshold, namely the $\psi$(3770) ($n$=1) and the $\psi$(4153) ($n$=2).
Both are $J^{PC}$=1$^{--}$ and broad, i.e.\ their experimentally measured width
is $>$20 MeV.
For the bottomonium system, there are two $D$-wave states predicted {\it below} threshold
and thus narrow, i.e.\ their hadronic width $\leq$30~keV \cite{upsilon1D_theory}.
Consequently, in case of an observation it might even be possible to resolve
the triplet of $J^{PC}$=0$^{--}$, 1$^{--}$ and 2$^{--}$.
An $\Upsilon$($1^3D_2$) candidate was observed by CLEO \cite{upsilon1D_cleo}
based upon 5.8$\times$10$^6$ $\Upsilon$(3S) decays with a mass
of 10161.1$\pm$0.6$\pm$1.6~MeV, consistent with a potential 
model prediction of 10158~MeV \cite{upsilon1D_potential}.

\noindent
BaBar searched for the $\Upsilon$($1^3D_2$) 
in a data set of 122$\times$10$^6$ $\Upsilon$(3S) decays \cite{upsilon1D_babar}.
The signal path has two radiative transitions 
$\Upsilon$(3S)$\rightarrow$$\Upsilon$(2P)$\gamma$ and 
$\Upsilon$(2P)$\rightarrow$$\Upsilon$(1D)$\gamma$, used for tagging,  
with subsequent decay by emission of a $\pi^+$$\pi^-$ pair to $\Upsilon$(1S),
which finally decays to $e^+$$e^-$ or $\mu^+$$\mu^-$.
The complete signal path was required for the identification.
BaBar was able to clearly confirm the $J^{PC}$=1$^{--}$ state
with 33.9$^{+8.2}_{-7.5}$ signal events (5.8$\sigma$ stat.\ significance).
The fitted mass is $m$=10164.5$\pm$0.8$\pm$0.5~MeV,
consistent with the CLEO measurement.

\section{Summary}

The $B$ factories continue to provide exciting results. Charmonium 
spectroscopy is studied in $B$ meson decays, $\gamma$$\gamma$ collisions 
and $\Upsilon$($n$S) decays. Bottomonium spectroscopy is studied in $\Upsilon$($n$S) decays. 
Highly excited states such as $L$=2 states are clearly identified and provide 
accurate tests for potential models. States which are not consistent 
with any potential model, such as the X(3872), are studied in new ways,
such as radiative decays or production in radiative decays.
Surprising properties such as large isospin violation are confirmed. 

\newpage

\begin{theacknowledgments}
We thank the KEKB group for excellent operation of the
accelerator, the KEK cryogenics group for efficient solenoid
operations, and the KEK computer group and
the NII for valuable computing and SINET3 network support.  
We acknowledge support from MEXT, JSPS and Nagoya's TLPRC (Japan);
ARC and DIISR (Australia); NSFC (China); MSMT (Czechia);
DST (India); MEST, NRF, NSDC of KISTI, and WCU (Korea); MNiSW (Poland); 
MES and RFAAE (Russia); ARRS (Slovenia); SNSF (Switzerland); 
NSC and MOE (Taiwan); and DOE (USA).
\end{theacknowledgments}

\bibliographystyle{aipproc}   
\bibliography{sample}

\IfFileExists{\jobname.bbl}{}
 {\typeout{}
  \typeout{******************************************}
  \typeout{** Please run "bibtex \jobname" to optain}
  \typeout{** the bibliography and then re-run LaTeX}
  \typeout{** twice to fix the references!}
  \typeout{******************************************}
  \typeout{}
 }


\end{document}